\documentclass[aps,pra,reprint, superscriptaddress,showpacs,longbibliography]{revtex4-1}
\usepackage[english]{babel}
\usepackage{amsmath}
\usepackage{amsfonts}
\usepackage{amssymb}
\usepackage{graphicx}
\usepackage{bm}
\usepackage{bbm}

\begin{document}
\title{Scalable gate architecture for densely packed semiconductor spin qubits}

\author{D. M. Zajac}
\affiliation{Department of Physics, Princeton University, Princeton, New Jersey 08544, USA}
\author{T. M. Hazard}
\affiliation{Department of Physics, Princeton University, Princeton, New Jersey 08544, USA}
\author{X. Mi}
\affiliation{Department of Physics, Princeton University, Princeton, New Jersey 08544, USA}
\author{E. Nielsen}
\affiliation{Sandia National Laboratories, Albuquerque, New Mexico 87185, USA}
\author{J. R. Petta}
\affiliation{Department of Physics, Princeton University, Princeton, New Jersey 08544, USA}


\begin{abstract}
We demonstrate a 12 quantum dot device fabricated on an undoped Si/SiGe heterostructure as a proof-of-concept for a scalable, linear gate architecture for semiconductor quantum dots. The device consists of 9 quantum dots in a linear array and 3 single quantum dot charge sensors. We show reproducible single quantum dot charging and orbital energies, with standard deviations less than 20\% relative to the mean across the 9 dot array. The single quantum dot charge sensors have a charge sensitivity of 8.2$\times$10$^{-4} e/\sqrt{\text{Hz}}$ and allow the investigation of real-time charge dynamics. As a demonstration of the versatility of this device, we use single-shot readout to measure a spin relaxation time $T_1$ = 170 ms at a magnetic field $B$ = 1 T. By reconfiguring the device, we form two capacitively coupled double quantum dots and extract a mutual charging energy of 200 $\mu$eV, which indicates that 50 GHz two-qubit gate operation speeds are feasible.
\end{abstract}

\maketitle

\section{Introduction}

The density of transistors in integrated circuits has been following Moore's law since its conception \cite{moore2005cramming}. However, as the size of transistors approaches the size of a single atom the laws of quantum physics will play an increasingly dominant role in computer architectures, making it difficult for this trend to continue much longer. Despite this, the prospect of utilizing quantum mechanical phenomena in information processing offers an opportunity to increase the computational power of computers, for some types of problems, beyond what is known to be possible on even the most ideal classical computer \cite{shor1999polynomial,grover1996fast}. In order for functional quantum computers to become a reality, they will require an on-chip physical component with reproducible properties that can be incorporated into large scale structures, much like the classical computer depends on the robustness of the transistor.

One of the leading candidates for the quantum analog of the transistor is the gate-defined, semiconductor quantum dot \cite{kouwenhoven1997,hanson2007spins}. The spin state of an electron trapped in a quantum dot is an ideal physical system for storing quantum information \cite{loss1998,divincenzo2000universal,taylor2005fault}. Silicon in particular, with its weak hyperfine fields, small spin-orbit coupling and lack of piezoelectric electron-phonon coupling, forms a ``semiconductor vacuum" for spin states \cite{steger2012}, and supports seconds-long electron spin coherence times  \cite{tyryshkin2012electron}. However, the fabrication of reliable and scalable Si based quantum dots has proved challenging. Independent of the need for a pure spin environment, quantum dots must have reproducible electrical properties for scaling. The relatively large effective mass of electrons in Si, along with the typically lower mobilities of Si two-dimensional electron gases, makes the fabrication of tightly confined, few-electron quantum dots with reproducible properties difficult \cite{payette2012single}.

\begin{figure*}
	\begin{center}
		\includegraphics[width=\textwidth]{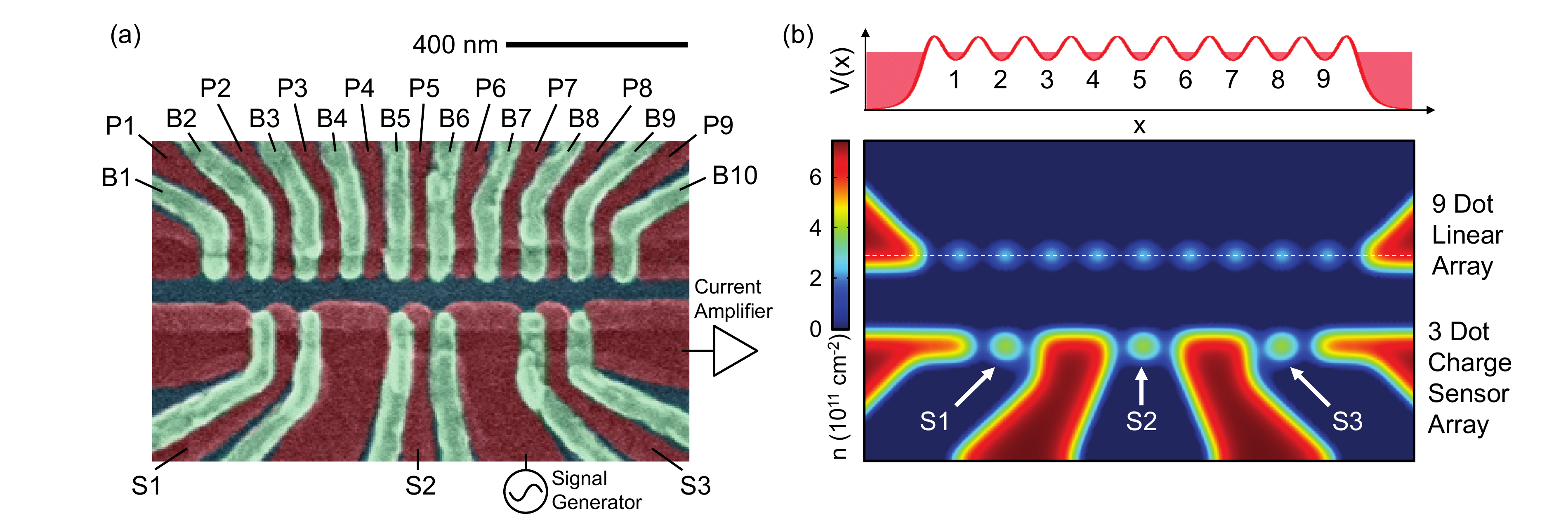}
		\caption{
			(a) False-color scanning electron microscope image of the overlapping gate architecture. A linear array of 9 quantum dots is formed under plunger gates P1, P2, ..., P9. Tunnel couplings are controlled using barrier gates B1, B2, ..., B10. Quantum dot charge sensors are formed under gates S1, S2, and S3. (b) Lower panel: COMSOL simulation of the electron density, $n$, in the quantum well. Upper panel: The confinement potential, $V(x)$, along the dashed line in the lower panel.
		}
		\label{fig1}
	\end{center}	
	\vspace{-0.6cm}
\end{figure*}

In this paper we present a path forward for scaling up semiconductor quantum dot devices. Our device consists of 12 quantum dots, 9 of which are arranged in a linear array, and 3 that are used as sensitive charge detectors. The enhancement mode device utilizes an overlapping aluminum gate architecture to achieve tight electronic confinement \cite{zajac2015reconfigurable}, while the undoped Si/SiGe heterostructure provides a clean, high mobility interface \cite{miPRB} for the formation of well-behaved quantum dots with reproducible characteristics. 

\section{Results and Discussion}

The outline of the paper is as follows. We first evaluate the reproducibility of the 9 dots in the array by extracting the critical parameters of single quantum dots formed under each plunger gate: the lever-arm, charging energy and orbital excited state energy. We are able to reach zero electron occupancy in all 9 quantum dots in the array, obtaining an average charging energy $E_\text{c}$ = 6.9 $\pm$ 0.7 meV and an average orbital energy $E_\text{orb}$ = 3.0 $\pm$ 0.5 meV. Using adjacent single quantum dots as charge sensors, we show that we are able to read out the charge state of the entire array with a signal-to-noise ratio (SNR) that allows for the observation of real-time tunneling events. Lastly, as a demonstration of the flexibility of the gate architecture, we perform single-shot spin readout and demonstrate strong capacitive coupling of two nearest-neighbor double quantum dots (DQDs).

\subsection{Linear Gate Architecture}

A false-colored SEM image of the device is shown in Fig.\ \ref{fig1}(a) and a COMSOL simulation of the electron density $n$ in the plane of the quantum well is shown in Fig.\ \ref{fig1}(b). Tight electronic confinement is achieved using an overlapping aluminum gate architecture \cite{zajac2015reconfigurable}. In the upper half of the device, two sets of aluminum gate electrodes, with a pitch of 100 nm, are interleaved to form a linear array of 9 quantum dots. A plunger gate controls the chemical potential of each quantum dot (shown in red), while barrier gates control the tunnel coupling of adjacent dots (shown in green). An aluminum screening layer restricts the action of the tuning gates to a one-dimensional channel \cite{zajac2015reconfigurable}. High sensitivity single electron charge detection is achieved using 3 single dot charge sensors defined in a second one-dimensional channel that is formed in the lower half of the device. 

The scalability of this device design is evident from its repeating unit cell structure. Each unit cell consists of 3 quantum dots and a charge sensor. The device demonstrated here is constructed by concatenating 3 of these unit cells. Scaling to arrays of arbitrary length is achievable by adding additional unit cells.  The overlapping gate architecture demonstrated here has roughly 4.5 times the areal density of a widely-used DQD depletion mode gate pattern; we fit 9 dots and 3 charge sensors in an area of $\sim1.5$ $\mu m^2$, the same area as a GaAs DQD and its two quantum point contact charge detectors \cite{petta2005}.

\subsection{Characterization of the 9 Dot Array}

Scaling to large arrays of quantum dots requires uniform and reliable single quantum dot characteristics. We adopt three figures of merit to characterize the reproducibility of the linear array: the lever-arm $\alpha$, charging energy $E_\text{c}$, and orbital excited state energy $E_\text{orb}$. We form a single quantum dot under each plunger gate with the neighboring quantum dots tuned to the many electron regime and extract $\alpha$, $E_\text{c}$ and $E_\text{orb}$ for each dot using a combination of transport measurements, charge sensing, and pulsed gate spectroscopy. 

\begin{figure}
	\includegraphics[width=0.95\columnwidth]{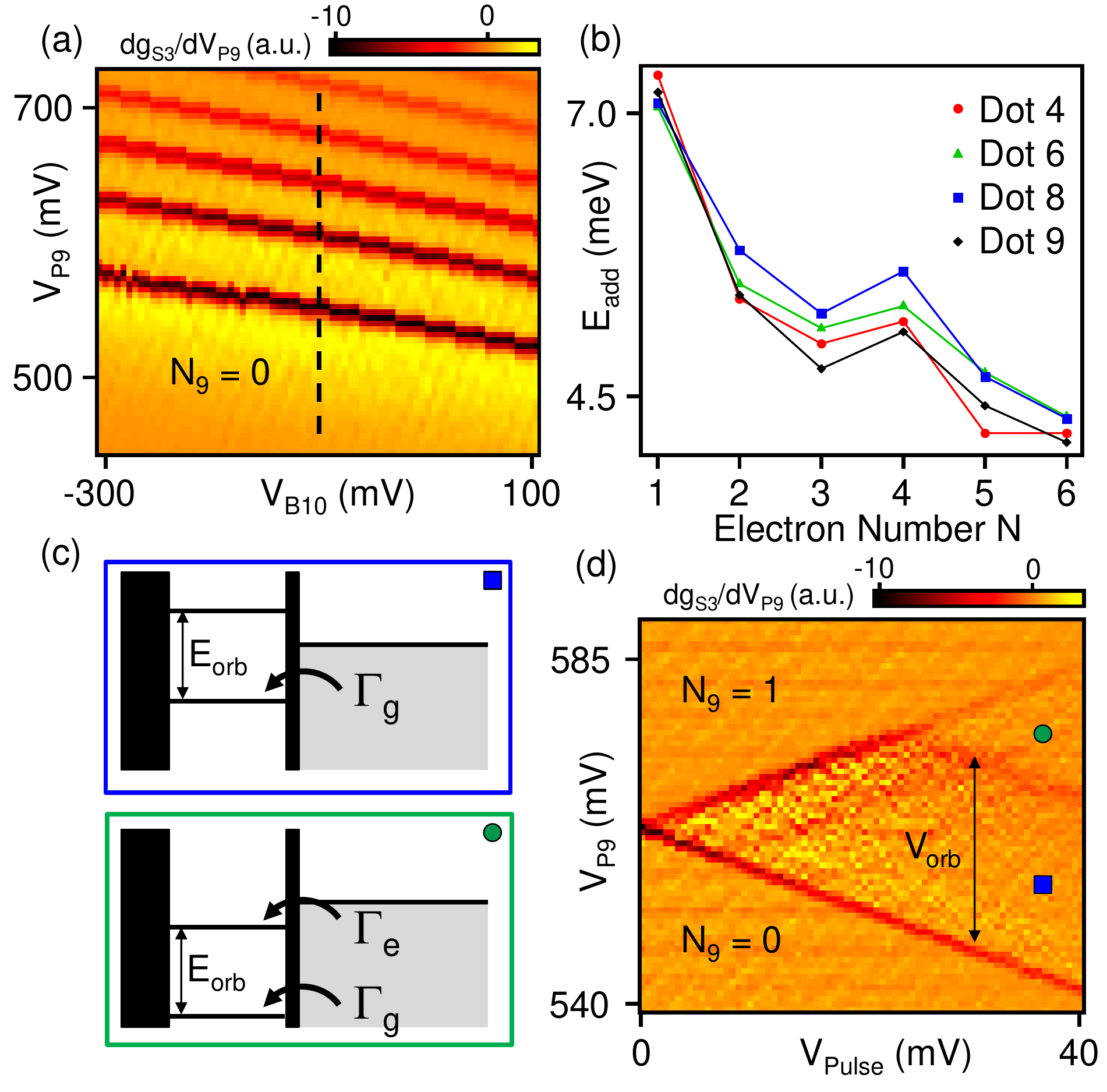}
	\caption{	
		(a) Charge stability diagram of quantum dot 9. The derivative of charge sensor dot 3 conductance, dg$_{\text{S3}}$/dV$_{\text{P9}}$, plotted as a function of V$_{\text{P9}}$ and V$_{\text{B10}}$. For low voltages, dot 9 is emptied of free electrons, reaching the $N_9$ = 0 charge state. (b) Addition energy, $E_{\rm add}$, plotted as a function of electron number $N$ for dots 4, 6, 8, and 9. (c) Pulsed gate spectroscopy: The effective tunneling rate onto the dot is dependent on $V_{\rm pulse}$. (d) An orbital excited state with energy $E_{\rm orb}$ = $\alpha V_{\rm orb}$ = 3.4 meV is visible in dot 9.
	}
	\label{fig2}
\end{figure}

Lever-arms are extracted from transport measurements of Coulomb diamonds at the $N$ = 0 to 1 transition, where $N$ refers to the number of electrons in the dot. The charge state of each dot is read out by measuring the conductance through the nearest single dot charge sensor. As an example, Fig.\ \ref{fig2}(a) shows the charge stability diagram of a quantum dot formed under plunger gate P9. Here the derivative of the charge sensor conductance, $dg_{\rm S3}/dV_{\rm P9}$, is plotted as a function of  $V_\text{P9}$ and  $V_\text{B10}$. The lack of charge transitions for low values of  $V_\text{P9}$ indicates that dot 9 has been emptied of free electrons, reaching $N_\text{9}$ = 0 charge occupancy. Addition voltages for dot 9 are extracted along the vertical dashed line in Fig.\ \ref{fig2}(a) and converted into addition energies, $E_\text{add}$, using $\alpha$. These addition energies are plotted in Fig.\ \ref{fig2}(b). For comparison, we also show the addition energies for dots 4, 6, and 8. The increase in $E_\text{add}$ at the $N$ = 4 to 5 charge transition is attributed to shell filling of the low lying spin and valley degrees of freedom \cite{yang2013spin,BorselliMagneto}. 

Pulsed gate spectroscopy is performed in each dot at the $N$ = 0 to 1 charge transition to extract the orbital excited state energy $E_\text{orb}$ \cite{elzerman2004excited,yang2012orbital}. A 500 Hz square wave with peak-to-peak amplitude $V_\text{pulse}$ is added to the dc plunger gate voltage to repeatedly load and unload an electron onto and off of the dot. For small $V_\text{pulse}$ only the ground state is pulled below the Fermi level of the lead [upper panel in Fig.\ \ref{fig2}(c)] and an electron tunnels onto the dot with a rate $\Gamma_\text{g}$. When the pulse amplitude exceeds $V_\text{orb}$, the electron can load into either the ground state or the first excited state [lower panel in Fig.\ \ref{fig2}(c)]. The effective loading rate is increased due to the contribution from the excited state $\Gamma_\text{e}$ and is visible in the charge sensing data as a change in the average sensor conductance. From these data we extract an orbital excited state energy $E_{\rm orb}$ = $\alpha V_{\rm orb}$ = 3.4 meV for dot 9.

Similar characterization is performed on dots 1--8 and the results are summarized in Table \ref{table1}. The averaged figures of merit are $\alpha$ = 0.13 $\pm$ 0.01 meV/mV, $E_\text{c}$ = 6.9 $\pm$ 0.7 meV, and $E_\text{orb}$ = 3.0 $\pm$ 0.5 meV. These charging energies are generally larger than those obtained with other device designs in Si/SiGe due to the tight confinement potential generated by the overlapping gate architecture. Specifically, depletion mode devices achieved charging energies of less than 2 meV \cite{yuan2011si}, while enhancement mode architectures have yielded charging energies close to 5 meV \cite{BorselliMagneto}. Moreover, the large orbital excited state energies are comparable to those measured in GaAs devices, where the effective mass is nearly three times smaller than Si \cite{hanson2007spins}.

\begin{table} [b]
	\begin{center}
		\begin{tabular}{c | c | c | c}
			Dot & $\alpha$ (meV/mV) & E$_\text{c}$ (meV)& E$_\text{orb}$ (meV)\\
			\hline\hline
			1 & 0.14 & 6.6 & 2.7\\
			2 & 0.13 & 6.1 & 2.6\\
			3 & 0.11 & 5.6 & 2.1\\
			4 & 0.14 & 7.3 & 3.3\\
			5 & 0.14 & 7.2 & 3.3\\
			6 & 0.14 & 7.1 & 3.0\\
			7 & 0.14 & 7.7 & 3.5\\
			8 & 0.14 & 7.1 & 3.4\\
			9 & 0.13 & 7.2 & 3.4\\
		\end{tabular}
		\caption{
			Lever-arm conversion between gate voltage and energy $\alpha$, charging energy $E_{\rm c}$, and orbital excited state energy $E_{\rm orb}$ for each of the 9 dots in the linear array.
		}
		\label{table1}
	\end{center}
\end{table}
\subsection{Sensitive Charge Detection}

\begin{figure}
	\begin{center}
		\includegraphics[width=0.95\columnwidth]{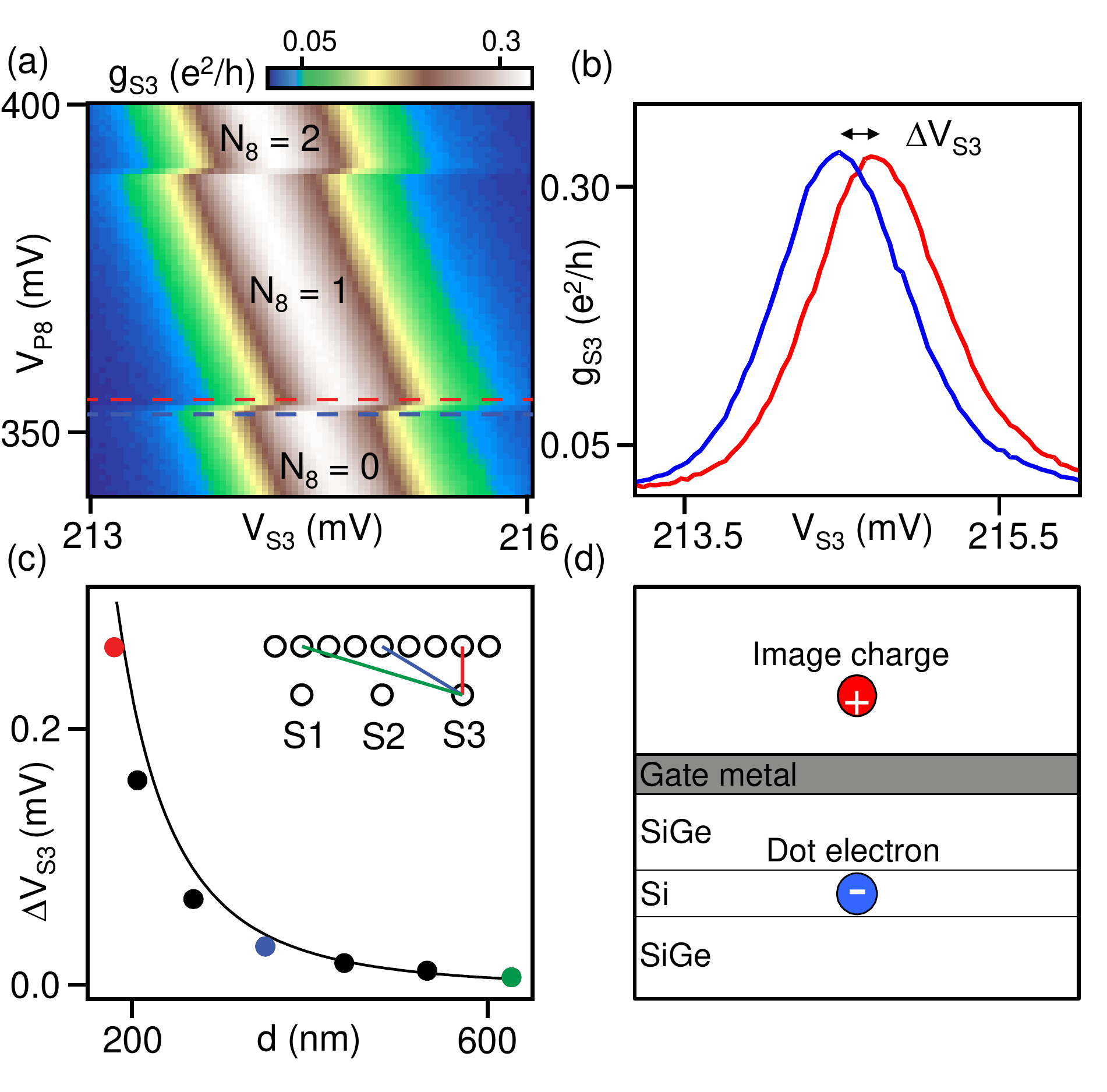}
		\caption{
			(a) A Coulomb blockade peak is visible in sensor dot 3 conductance, $g_{\rm S3}$,  which is plotted as a function of the gate voltages $V_{\rm P8}$ and $V_{\rm S3}$. (b) $g_{\rm S3}$ measured at the locations indicated by the dashed lines in panel (a). The Coulomb blockade peak shifts by $\Delta V_{\rm S3}$ = 0.26 mV when $N_8$ changes by one electron. (c) $\Delta V_{\rm S3}$ is measured for dots 2--8 and plotted as a function of the distance $d$ from the sensor dot. The black line is the theoretical prediction. (d) The approximately $1/d^3$ power law dependence is qualitatively understood as the field of a dipole formed by the electron in the quantum well (blue circle) and its positive image charge (red circle).
		}
		\label{Fig3}
	\end{center}
	\vspace{-0.6cm}
\end{figure}

An important criterion for quantum information processing is high fidelity qubit readout. For both single-shot readout of an individual spin \cite{elzerman2004single,morello2010} and spin-to-charge conversion in double \cite{petta2005} and triple quantum dot qubits \cite{medford2013} this translates to a need for high fidelity charge state readout. We demonstrate high sensitivity charge detection using the charge sensor array. The 3 sensor dots give good coverage over the entire 9 dot array.

In order to characterize the charge sensor performance we first measure the shift in a charge sensor Coulomb blockade peak due to a change in the charge occupancy of a nearby dot in the linear array. As an example, in Fig.\ \ref{Fig3}(a), we plot the conductance through charge sensor 3, $g_\text{S3}$, as a function of $V_\text{P8}$ and $V_\text{S3}$. A Coulomb blockade peak is visible in the sensor dot conductance and it abruptly shifts each time an electron is added to quantum dot 8. We measure a peak shift of $\Delta V_\text{S3}$= 0.26 mV at the $N_8$ = 0 to 1 charge transition. The shift in the charge sensor 3 Coulomb blockade peak position is also measured for dots 2--7 and is plotted in Fig.\ \ref{Fig3}(c) as a function of the geometric distance, $d$, between each dot and the sensor dot. The shift falls off with a power law that is close to $1/d^3$.

Predictions for the shifts in the sensor dot Coulomb blockade peak position can be obtained by computing the capacitances of the device. We construct a 3-dimensional model of the device based on the wafer growth profile and lithographic gate dimensions, representing the dots as metallic cylinders with a radius of 19 nm and height of 5 nm, each centered 7 nm below the Si/SiGe interface. The capacitances of the device are then computed using the fast-multipole-moment solver FastCap \cite{fastcap}. The expected shift is computed from the simulated capacitances using $\Delta V_{\rm S3} = \frac{e C_{\rm m}}{C_{\rm p}C_{\rm t}}$ where $C_{\rm m}$ is the mutual capacitance between the sensor dot and the single-electron dot, $C_{\rm p}$ is the capacitance between the sensor dot and its plunger gate, and $C_{\rm t}$ is the total single-electron dot capacitance \cite{van2002electron}. The computed shift scales as $\Delta V_{\rm S3} (d) \propto 1/d^{3.02 \pm 0.05}$ and agrees nicely with the experimental data [see the solid black line in Fig.\ \ref{Fig3}(c)]. We point out that the accuracy of this model is limited by uncertainty in the exact location of the quantum dots in the quantum well.

As in the case of a parallel plate capacitor, one might expect the capacitance to scale as 1/$d$. However, the overlapping gate architecture covers nearly the entire Si/SiGe heterostructure with metal, resulting in a significant amount of screening. The impact of this screening can be understood using the method of images charges [Fig.\ \ref{Fig3}(d)]. An electron trapped in a quantum dot induces a positive image charge in the gate metal above. The resulting electric field due to the electron and its image charge is that of a dipole, which falls off with a 1/$d^3$ dependence.

\subsection{Real-Time Charge Detection}

The ability to resolve real-time charge dynamics allows the study of fundamental physical phenomena at the level of single electrons \cite{kung2012irreversibility,maisi2016prl}. It also enables single-shot readout of single electron spin states \cite{elzerman2004single,morello2010} and the discrimination of two-electron singlet and triplet spin states \cite{petta2005}. We now demonstrate high sensitivity charge detection through the observation of real-time tunneling events \cite{vandersypen2004real,gustavsson2009electron}. Through a quantitative analysis of the charge sensor response, we extract a charge sensitivity of 8.2$\times$10$^{-4} e/\sqrt{\text{Hz}}$.

\begin{figure}
	\begin{center}
		\includegraphics[width=\columnwidth]{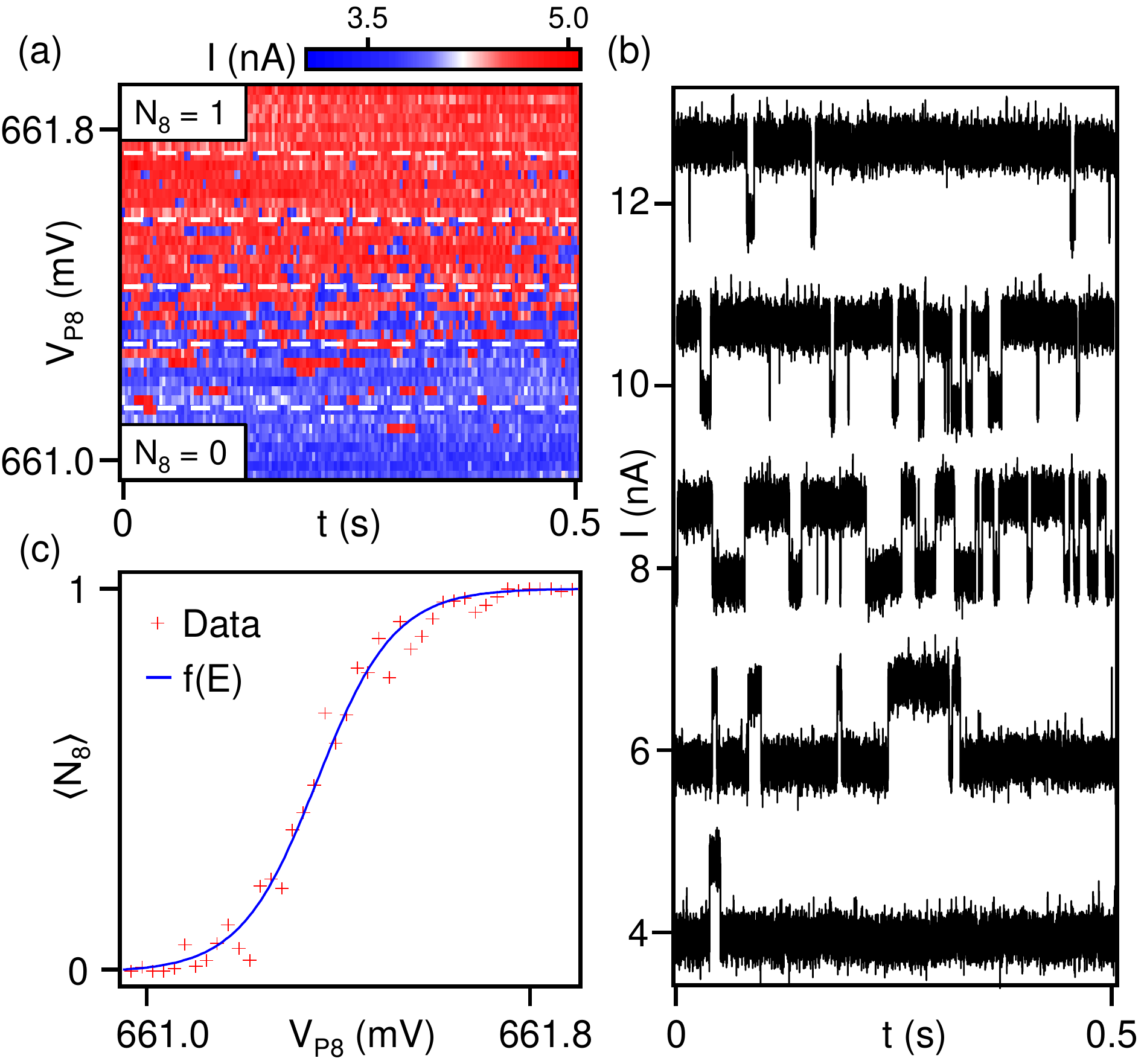}
		\caption{
			(a) The current, $I$, through sensor dot 3, plotted as a function of $V_{\rm P8}$ and time, $t$, near the $N_8$ = 0 to 1 charge transition. (b) Time series extracted from the data in (a) at the positions shown by the dashed lines. The dwell time in the $N_8$ = 1 charge state increases as $V_{\rm P8}$ is made more positive. The traces are offset by 2 nA for clarity. (c) Time-averaged quantum dot 8 occupation, $\langle N_{\rm 8} \rangle$, extracted from the data in (a) and plotted as a function of $V_{\rm P8}$. The data are fit to a Fermi function $f(E)$.
		}
		\label{Fig4}
	\end{center}
	\vspace{-0.6cm}
\end{figure}

Figure \ref{Fig4}(a) shows a color-scale plot of the current $I$ through sensor dot 3 as a function of time, for a range of plunger gate voltages $V_\text{P8}$ with dot 8 tuned up near the $N_\text{8}$ = 0 to 1 charge transition. Five time series extracted from this data set are plotted in Fig.\ \ref{Fig4}(b). The lowest time-series in Fig.\ \ref{Fig4}(b) was acquired with $V_\text{P8}$ = 661.12 mV. Here the dot is empty nearly all of the time. With $V_\text{P8}$ slightly increased, the current shows signatures of real-time single electron tunneling events and switches between two levels corresponding to the $N_8$ = 0 and 1 charge states. As expected, the dwell time in the $N_\text{8}$ = 1 charge state increases with increasing $V_\text{P8}$. Using a threshold to discriminate between the charge states, we plot the time-averaged occupation of dot 8, $\langle N_\text{8} \rangle$, as a function of $V_\text{P8}$ in Fig.\ \ref{Fig4}(c). We expect the population to follow a Fermi function as the chemical potential of the dot level is lowered past the Fermi level of the lead. The data in Fig.\ \ref{Fig4}(c) are nicely fit to a Fermi function with an electron temperature $T_{\rm e}$ = 120 mK.

\begin{figure}[t]
	\begin{center}
		\includegraphics[width=\columnwidth]{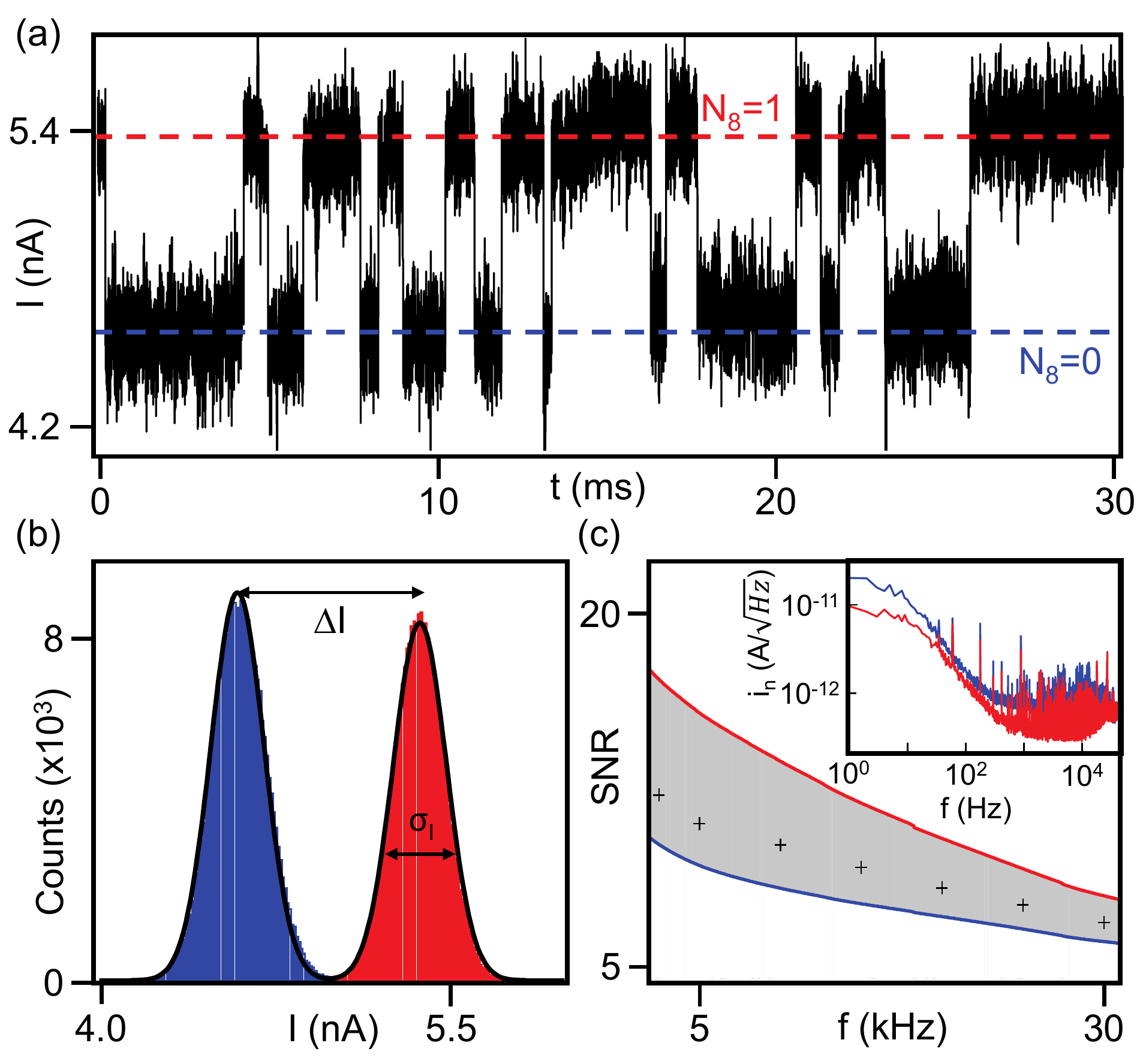}
		\caption{	
(a) A time series of the current, $I$, through sensor dot 3, with dot 8 configured at the $N_{\rm 8}$ = 0 to 1 charge transition. (b) A histogram of a one second time series exhibits two Gaussian peaks with width $\sigma_{\rm I}$ = 0.112 nA separated by an amount $\Delta I$ = 0.772 nA. (c) The SNR = $\Delta I/\sigma_{\rm I}$ plotted as a function of the filter cutoff frequency, $f$ (black crosses). The data fall between the expected SNR for a current level of 4 nA (blue) and 6 nA (red) based on the measured noise spectra at each current level, shown in the inset.			
}
		\label{Fig5}
	\end{center}
	\vspace{-0.6cm}
\end{figure}

A detailed analysis of the real-time single electron tunneling events can be used to determine the charge sensor SNR and charge sensitivity. We first measure a one-second time series of the current through the charge sensor with dot 8 tuned to the $N_\text{8}$ = 0 to 1 charge degeneracy point. The data are acquired at a sampling rate of 500 kHz and a Kaiser-Bessel finite impulse response (FIR) filter is used to reduce the effective measurement bandwidth to 30 kHz, the 3 dB point of our room temperature amplifier. A 30 ms long segment of this time series is shown in Fig.\ \ref{Fig5}(a). Real-time tunneling events between $N_\text{8}$ = 0 and $N_\text{8}$ = 1 are seen as two level switching in the measured current. A histogram of the full time trace is shown in Fig.\ \ref{Fig5}(b). The two well-resolved peaks correspond to the two charge states. Each peak is nicely fit to a Gaussian with width $\sigma_\text{I}$ = 0.112 nA, corresponding to the current noise in our measurement setup. The centroids of the two Gaussians are separated by $\Delta I$ = 0.772 nA, which corresponds to the signal associated with a  change in electron occupancy of one. For these data we extract a SNR = $\Delta I$/$\sigma_\text{I}$ = 6.9. By adjusting the FIR filter cutoff frequency, $f$, we plot the SNR as a function of the effective measurement bandwidth in Fig.\ \ref{Fig5}(c), showing a decrease in the SNR with increasing $f$.

A quantitative description of the SNR requires a more careful analysis of the experimental setup. We therefore measure the current noise of the device. The measured noise spectra, $i_{\rm n}(f)$, at current levels of 4 nA (6 nA) are plotted as the blue (red) traces in the inset of Fig.\ \ref{Fig5}(c). The noise is approximately white at high frequencies, but noise with an approximate $1/f$ dependence dominates at low frequency, and the overall noise level appears to be correlated to the derivative of the charge sensor current with respect to gate voltage. We can use these spectra to calculate the expected noise for a one second time series by integrating over frequency from 1 Hz to the filter cutoff frequency, $f$:
\begin{equation}
\sigma^2_\text{I}(f)=\int\limits_{1 \text{Hz}}^{f} i^2_n(f') df'.
\end{equation}

Using the measured signal $\Delta I$ = 0.772 nA, we plot the expected SNR as a function of $f$ in Fig.\ \ref{Fig5}(c). The measured SNR falls within the shaded region between the two curves that delineate the expected SNR for current levels of 4 and 6 nA. For a 30 kHz bandwidth the SNR = 6.9, implying an effective charge sensitivity of  8.2$\times$10$^{-4} e/\sqrt{\text{Hz}}$. This sensitivity is higher than both the rf-QPC ($\sim$ $10^{-3}$ $e/\sqrt{\text{Hz}}$)   \cite{reilly2007fast} and dispersive gate readout (6.3 $\times$ $10^{-3}$ $e/\sqrt{\text{Hz}}$)  \cite{colless2013dispersive}, however our measurement bandwidth is limited to 30 kHz due to our current amplifier. Improvements to the SNR and measurement bandwidth could be made by using a low temperature preamplifier \cite{curry2015cryogenic} in combination with a higher bandwidth room temperature amplifier.

\subsection{Versatility Demonstrations}

The 9 dot linear array is capable of hosting a diverse range of quantum dot qubits. Using individual spins, 9 nearest-neighbor exchange coupled Loss-DiVincenzo qubits can be formed within the array \cite{loss1998}. With the gate voltages configured differently, four singlet-triplet qubits could be formed using pairs of electrons \cite{petta2005} and the qubits could be coupled via a dipole-dipole interaction \cite{Shulman2012}. Alternatively, three exchange-only spin qubits could be defined, allowing full electrical control over the  Bloch sphere of each qubit \cite{medford2013,medford2013self,eng2015isotopically}. To demonstrate the versatility of this device architecture we first perform single-shot readout of an electron spin to measure the spin lifetime, $T_1$. Finally, we form two capacitively coupled DQDs and measure an interaction strength of 200 $\mu$eV, which suggests a 50 GHz two-qubit gate operation speed.

\begin{figure}[t]
	\begin{center}
		\includegraphics[width=\columnwidth]{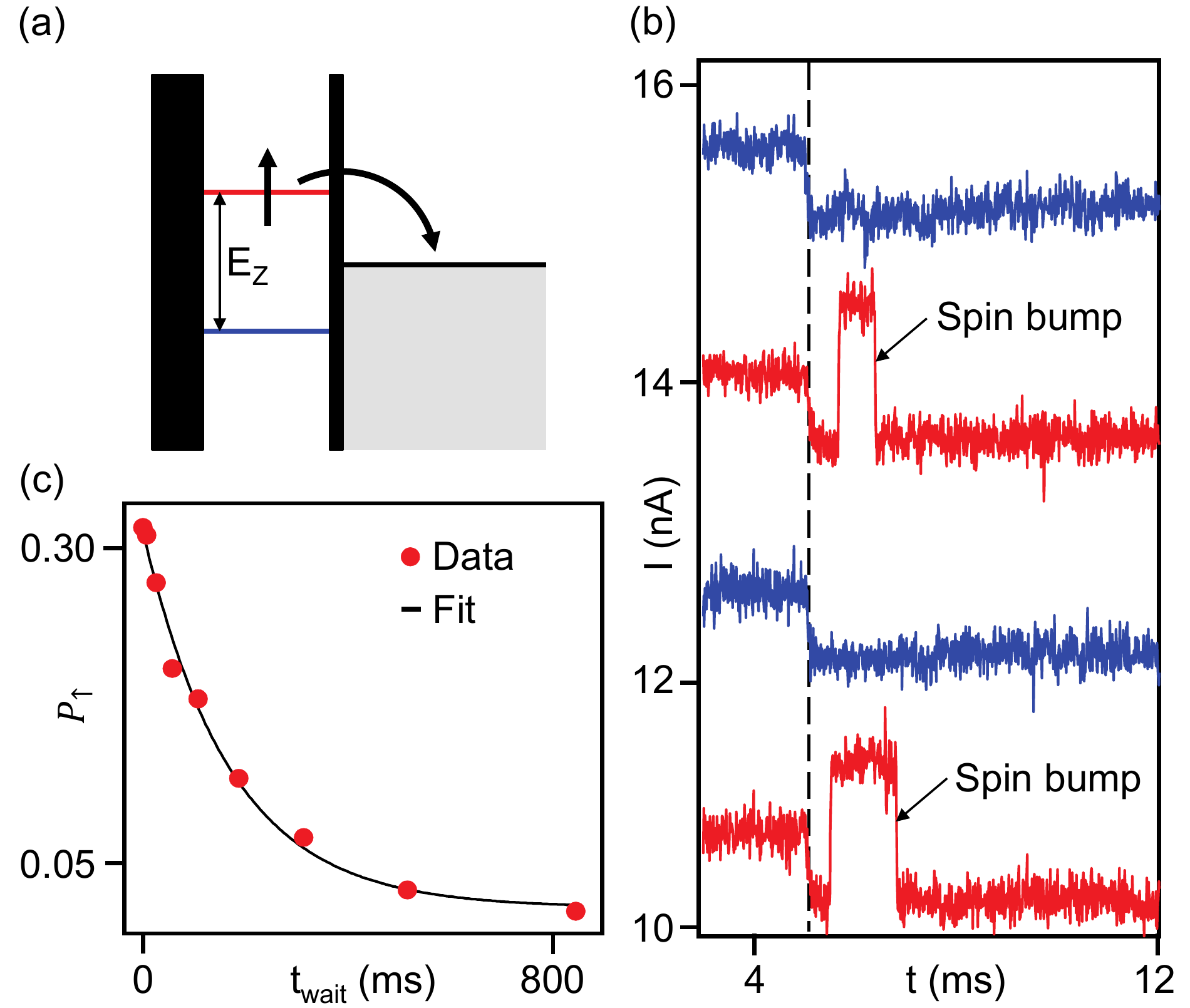}
		\caption{
		Single-shot spin measurements. (a) A single electron spin is measured by aligning the spin states such that a spin up electron (red level) can tunnel off of the dot and be replaced by a spin down electron, while a spin down electron (blue level) does not have sufficient energy to tunnel off of the dot. (b) Example single-shot traces acquired at $B$ = 1 T. The vertical dashed line at $t$ = 5 ms marks the beginning of the readout phase. Red (blue) traces correspond to spin up (spin down) electrons. Spin up events result in a ``spin bump." (c) $P_{\uparrow}$ decays exponentially with $t_{\rm wait}$, with a best fit $T_1$ = 170 $\pm$ 17 ms.
		}
		\label{Fig6}
	\end{center}
	\vspace{-0.8cm}
\end{figure}

We now demonstrate single-shot spin state readout on dot 8 in the linear array. A three-step pulse sequence is employed to measure the spin relaxation time $T_1$ at a magnetic field $B$ = 1 T \cite{elzerman2004single,morello2010}. Starting with an empty dot, we plunge the chemical potential of the dot level far below the Fermi level of the lead, which allows an electron to load into either the spin up or the spin down state. After a time $t_\text{wait}$ we begin the readout phase by setting the chemical potential of the dot such that the spin up and spin down energy levels straddle the Fermi level of the lead. If the electron on the dot is in the spin-up excited state, as shown in Fig.\ \ref{Fig6}(a), the electron will hop off of the dot and then be replaced by a spin-down electron. The change in the charge occupancy of the quantum dot due to this process is visible in time series measurements of the sensor dot current, $I$, and is referred to here as a ``spin bump." In contrast, if the final spin state is spin down, no spin bump will be observed.  Lastly, we raise the chemical potential of both spin states above the Fermi level to empty the dot and complete the measurement cycle.

Example single-shot traces are shown in Fig.\ \ref{Fig6}(b). Spin up electrons are indicated by current pulses during the readout phase (red traces) while spin down electrons simply remain on the dot during the readout phase (blue traces). We extract $T_1$ by varying $t_\text{wait}$ and measuring the probability $P_{\uparrow}$ of being in the spin up state at the end of the measurement phase [see Fig.\ \ref{Fig6}(c)]. Each data point represents the average of 10,000 single-shot traces. The resulting data are fit to an exponential decay with a best fit $T_1$ = 170 $\pm$ 17 ms. The long spin relaxation time is a defining feature of the Si ``semiconductor vacuum."

\begin{figure}[t]
	\begin{center}
		\includegraphics[width=\columnwidth]{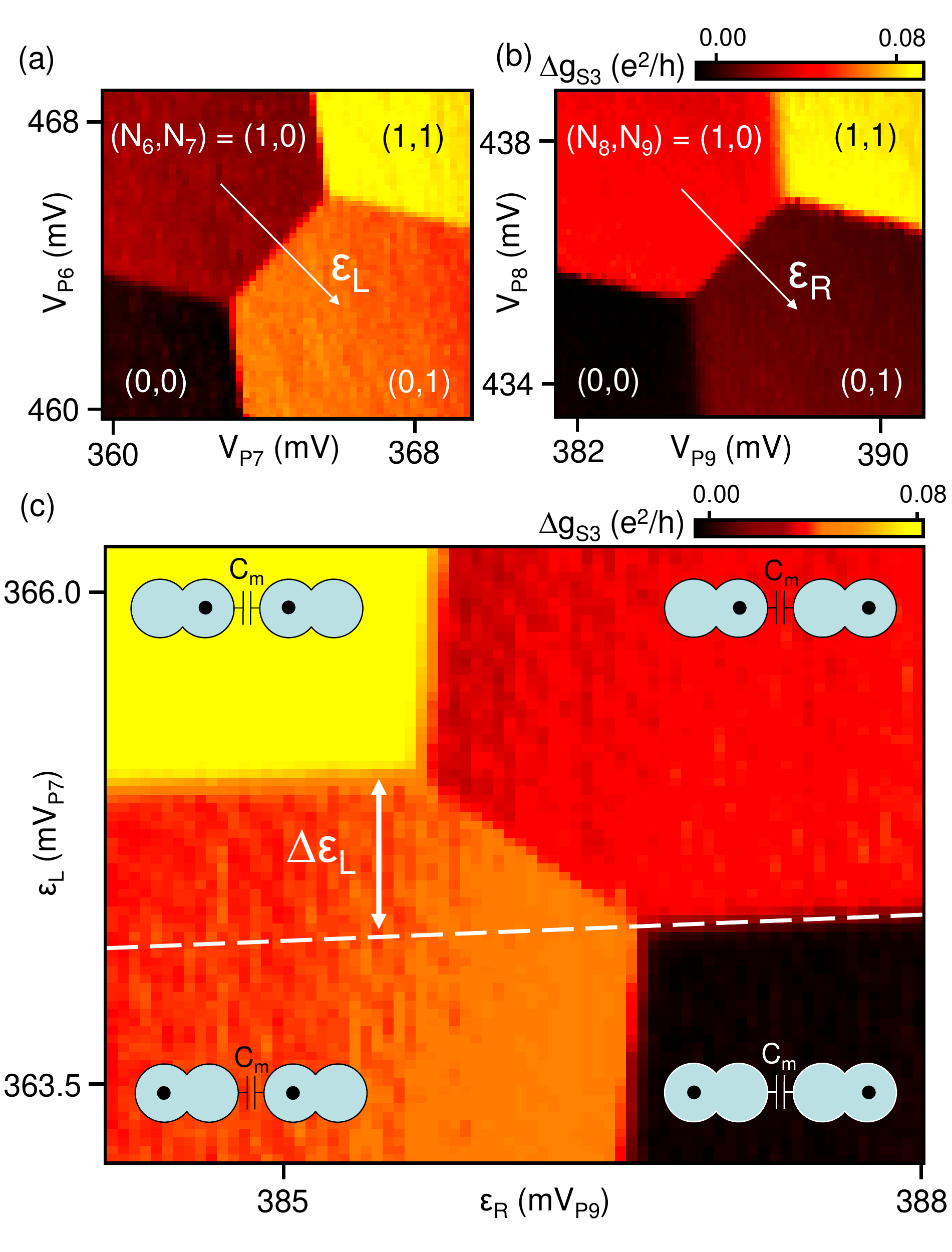}
		\caption{
			Dots 6--7 and 8--9 are simultaneously tuned up to form two DQDs, as shown by the charge stability diagrams in panels (a--b). (c) The capacitive interaction between the two DQDs is extracted by measuring the quadruple dot charge stability diagram as function of $\varepsilon_{\text{L}}$ and $\varepsilon_{\text{R}}$. The ($N_6$, $N_7$) = (1,0) to (0,1) interdot charge transition shifts by $\Delta \varepsilon_{\rm L}$ = 0.77 mV when $\varepsilon_{\rm R}$ is swept across the ($N_8$, $N_9$) = (1,0) to (0,1) interdot charge transition.
		}
		\label{Fig7}
	\end{center}
	\vspace{-0.8cm}
\end{figure}

Capacitive coupling has been proposed to mediate two-qubit interactions \cite{taylor2005fault}. Our compact gate design leads to large capacitive couplings. As a demonstration, we investigate the capacitive coupling of two adjacent DQDs. We use dots 6--7 to define one DQD and dots 8--9 to define a second DQD. The charge stability diagrams for these DQDs are shown in Figs.\ \ref{Fig7}(a--b). The barrier gate voltage $V_{\rm B8}$ is set such that there is no tunneling between dots 7 and 8. As a result, the two DQDs are coupled only via a capacitive interaction $C_{\rm m}$. Interdot detuning axes, $\varepsilon_\text{L}$ and $\varepsilon_\text{R}$, are overlaid on the data in Figs.\ \ref{Fig7}(a--b). By sweeping $\varepsilon_\text{L}$ vs $\varepsilon_\text{R}$, we obtain the quadruple quantum dot stability diagram shown in Fig.\ \ref{Fig7}(c). The mutual capacitance $C_{\rm m}$
causes the ($N_6$, $N_7$) = (1,0) to (0,1) interdot charge transition to shift by $\Delta \varepsilon_{\rm L}$ = 0.77 mV when the occupancy of the second DQD changes from ($N_8$, $N_9$) = (1,0) to (0,1).  Using the lever-arm conversion between gate voltage and energy, this corresponds to a 200 $\mu$eV energy shift (50 GHz two-qubit gate operation time) \cite{ward2016state}.

\section{Conclusion}
In summary, we have developed a scalable quantum dot gate architecture that yields quantum dots with uniform and reproducible characteristics. As a proof-of-concept, we have presented a 12 quantum dot device consisting of a linear array of 9 quantum dots and 3 single quantum dot charge sensors. From characterization measurements we obtain standard deviations in the charging energies and orbital energies of less than 20\% relative to their means: $E_\text{c}$ = 6.9 $\pm$ 0.7 meV, $E_\text{orb}$ = 3.0 $\pm$ 0.5 meV. We have demonstrated the ability to detect real-time tunneling events in this large array, and used this capability for single-shot measurements of the electron spin. As a final demonstration, we characterize the dipole-dipole coupling of two adjacent DQDs formed in the array and measure an interaction energy of 200 $\mu$eV, which bodes well for computing architectures that rely on capacitive coupling of qubits.

\section{Acknowledgments}

 Funded by the ARO through grant No.\ W911NF-15-1-0149, the Gordon and Betty Moore Foundation's EPiQS Initiative through Grant GBMF4535, and the NSF (DMR-1409556 and DMR-1420541). Devices were fabricated in the Princeton University Quantum Device Nanofabrication Laboratory.

\bibliographystyle{apsrev4-1}
\bibliography{citationsPhysRevApp2}

\end{document}